\documentclass[a4paper,11pt,twoside]{scrartcl}
\usepackage{ILD}

\usepackage{xspace}

\usepackage{xpatch}
\makeatletter
\xpatchcmd\@HepConStyle
 {\edef\@upcode{\updefault}}
 {\ifdefined\shapedefault\edef\@upcode{\shapedefault}\else\edef\@upcode{\updefault}\fi}
 {}{}
\makeatother

\title{Searches
  for the leptophilic Z' boson at the International Linear Collider
  and Linear Collider Facility}

\ildproc{PHYS}{2026}{001}

\date{March 17, 2026}

\addauthor{Aleksander Filip \.Zarnecki}{\institute{1}}

\addinstitute{1}{Faculty of Physics, University of Warsaw, Poland\\
                          filip.zarnecki@fuw.edu.pl}

\abstract{
  Production of the leptophilic Z' boson at future colliders
  was proposed as one of the benchmark Beyond the Standard Model (BSM)
  scenarios for the Physics Briefing Book prepared for the 2026 update
  of the European Strategy for Particle Physics. Presented in this
  contributions are results on the sensitivity of the experiments at
  the International Linear Collider (ILC) and at the Linear Collider
  Facility (LCF) at CERN to the considered scenario. Studied as a
  function of the Z' boson mass is the expected signal in the di-muon
  decay channel, which is expected to have the highest
  sensitivity. Signal and background simulation in \textsc{Whizard}
  includes effects of multi-photon emission and initial state
  radiation (ISR), with proper matching between ISR and matrix element
  domains. Expected limits on the Z' couplings to SM leptons are
  presented for ILC and LCF running at 250\,GeV and LCF running at
  550\,GeV.
}

\titlecomment{This work was carried out in the framework of the ILD Concept Group} 

\newcommand{\ee}{\ensuremath{\textrm{e}^{_+}\!\textrm{e}^{_-}}\xspace}
\newcommand{\mumu}{\ensuremath{\mu^{_+}\!\mu^{_-}}\xspace}
\newcommand{\mumug}{\ensuremath{\mu^{_+}\!\mu^{_-} \gamma}\xspace}

\newcommand{\abinv}{\ensuremath{\text{ab}^{-1}}\xspace}

\newcommand{\whizard}{\textsc{Whizard}\xspace}

\newcommand{\delphes}{\textsc{Delphes}\xspace}

\graphicspath{{./logos/} {./plots/} }

\begin{document}


\titlepage
\pagenumbering{arabic}\setcounter{page}{2}

\section{Motivation}

We are all convinced that the Standard Model (SM) can not be the
ultimate theory describing the universe. It does not include
gravity and fails to describe many key observations like the existence
of dark matter (DM), the matter-antimatter asymmetry in the Universe or
neutrino oscillations.
From pletora of the BSM models proposed (and not yet excluded by the
LHC data) many include new interactions mediated by the new gauge
bosons, and the corresponding collider phenomenology strongly depends
on their couplings to the SM fields.
One of the interesting scenarios proposed at the Open Symposium in
Venice  \cite{RGS:Venice} as the BSM benchmark
to be studied for different collider options in the recent update of
the European Strategy for Particle Physics (ESPP) is a model with a
leptophilic Z' boson, assuming the $L_{e} \!-\! L_{\mu}$ U(1) gauge
group and a vector Z' coupling to SM leptons.
Presented in this contribution are results of the rapid analysis
initiated after the Venice symposium for ILC and LCF running at
250\,GeV and then extended to LCF running at 550\,GeV.
Results for 250\,GeV were included in the ESPP Physics Briefing Book~\cite{deBlas:2025gyz}.

\section{Model}

Considered in this study is a model with a leptophilic Z' boson, with vector
coupling to electrons and muons only (and the corresponding neutrinos). 
While avoiding gauge anomaly and limits from Electroweak Precision
Observables (EWPO), this scenario is very challenging for hadron
colliders, where only associated production can be considered, and the
projected HL-LHC bounds are not expected to be stronger than the
existing LEP2 limits, which are also relatively weak.
On the other hand, future lepton colliders offer high sensitivity for
Z' searches in multiple channels \cite{GonzalezSuarez:2024dsp,Yue:2024kwo}. 

No dedicated studies addressed prospects for leptophilic Z' searches
at the ILC before.
However, coupling limits from the search for DM production by exchange
of a light mediator with vector couplings~\cite{Kalinowski:2021tyr}
can be also translated to the limits on the Z' couplings to leptons in the
considered model.
This is because one third of Z' decays is expected to be invisible
(decays to SM neutrino pairs) and result in mono-photon signature (as
for the light DM pair-production).
Still, sensitivity in the invisible channel is limited.
As shown in an FCC-ee study~\cite{GonzalezSuarez:2024dsp},
the best sensitivity to light Z' production is expected in the \mumu
decay channel.
This is why the di-muon channel was selected for a dedicated ILC/LCF
sensitivity study. 

\section{Event simulation}

Event samples for the presented analysis were generated with
\whizard~3.1.6 \cite{Kilian:2007gr,Moretti:2001zz} and processed with
\delphes \cite{deFavereau:2013fsa} using \texttt{ILCgen} model for
fast detector simulation and event reconstruction. 
Signal of light Z' boson production, for $m_{Z'} < \sqrt{s}$, is
dominated by on-shell production in the radiative return \ee annihilation.
For the considered \mumu decay channel, 
the leading-order (LO) process is
\[ \ee \to Z' \, \gamma   \to \mumug\;. \]
Neglecting the beam energy spread, mono-energetic Z' (di-muons) and
photons are expected in the LO final state, with energies given by
\begin{eqnarray*}
  E_{Z'}  =  \frac{\sqrt{s}}{2} +
  \frac{M_{Z'}^{2}}{2\,\sqrt{s}} & ~~~\text{and}~~~ &
   E_{\gamma}  =  \frac{\sqrt{s}}{2} -
   \frac{M_{Z'}^{2}}{2\,\sqrt{s}}
\end{eqnarray*}
However, this is no longer the case when initial state radiation (ISR)
and processes with multi-photon emission are taken into account.
For a proper description of the event kinematics we follow the
approach described in~\cite{Kalinowski:2020lhp}:
we generate event in \whizard with up to three photons on the matrix element (ME),
include effects of soft and collinear ISR emissions with lepton ISR
structure functions and apply proper ISR-ME matching to avoid double-counting.
Including these effects is important, as they result in up to
\SI{40}{\%} increase in the signal cross section compared to the LO
(single photon emission) approximation without ISR.
\begin{figure}[tbp]
    \centering
    \includegraphics[width=0.4\textwidth]{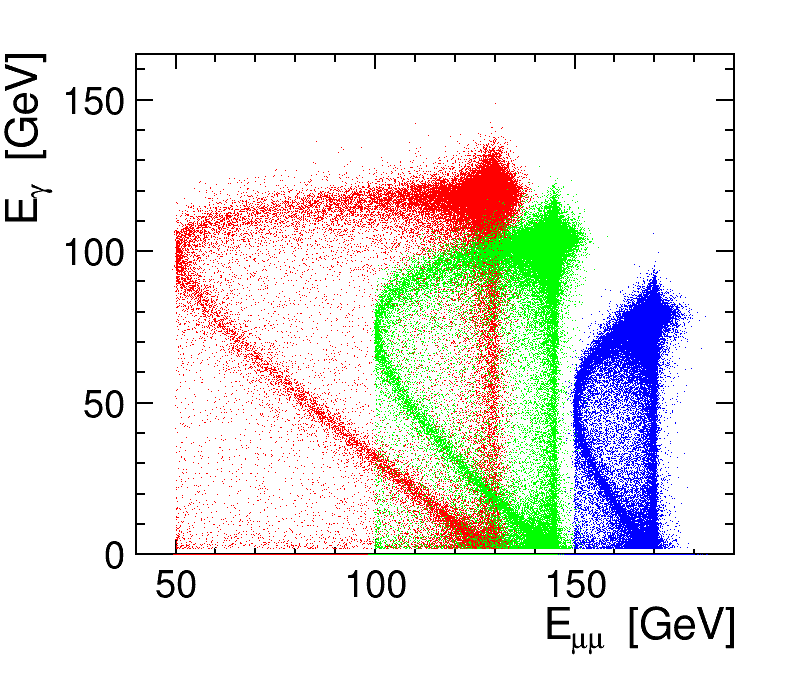}
    \includegraphics[width=0.45\textwidth]{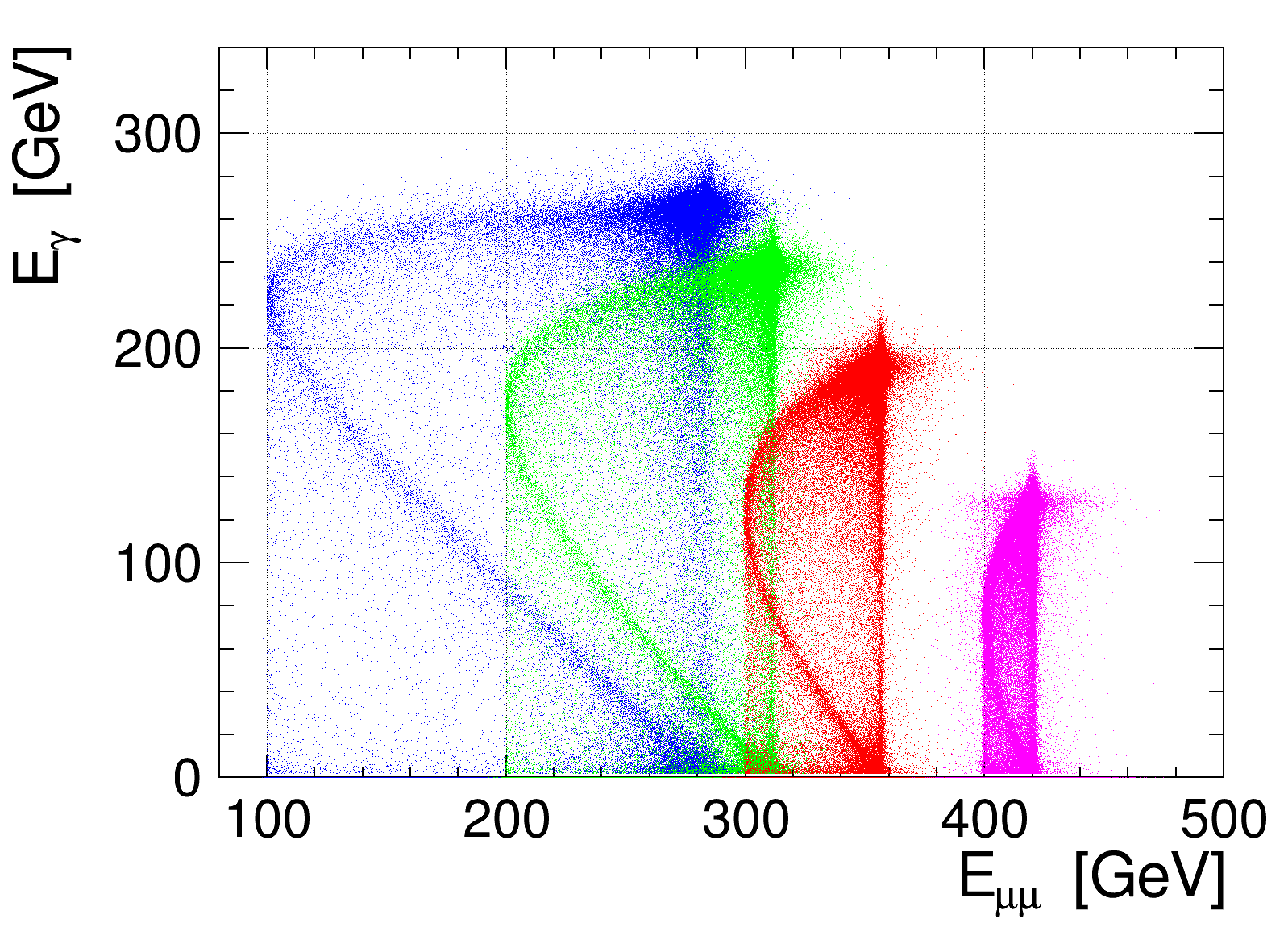}
    \caption{Distribution of the energy for the muon pair and for the
      highest energy photon in Z' production events simulated with
      \whizard.
      Left: for ILC running at 250\,GeV and Z' mass of 50\,GeV (red),
      100\,GeV (green) and 150\,GeV (blue).
      Right: for LCF running at 550\,GeV and Z' mass of 100\,GeV (blue),
      200\,GeV (green), 300\,GeV (red) and 400\,GeV (magenta).
    }  
    \label{fig:energy}
\end{figure}
Higher order effects also result in significant distortions of
the reconstructed energy distributions for the di-muon system (Z'
candidate) and the radiated photons, as shown  in Fig.~\ref{fig:energy}.

Based on the considered signal signature --- \mumu or \mumug final state
reconstructed in the detector --- only processes with leptonic final
states (and possibly radiated photons) were included in the simulated SM background.
Considered was radiative muon and taon pair-production,
$\mu^{_+}\!\mu^{_-} (\gamma)$ and $\tau^{_+}\!\tau^{_-} (\gamma)$, as
well as four-fermion processes with possible photon emission:
$\mu^{_+}\!\mu^{_-} \ell^{_+}\!\ell^{_-}(\gamma)$ and
$\mu^{_+}\!\mu^{_-} \nu\bar{\nu} (\gamma)$.
Possible muon pair production in processes involving quasi-real
photons, as described by the Equivalent Photon Approximation, was also
included.
For each signal and background channel, samples for each beam
polarisation combination (assuming 100\% polarised beams) were
generated separately and then mixed together with proper weights.
Selected for final analysis were events with only two muons and
arbitrary number of photons reconstructed in the final state.

\section{Event selection}

For majority of radiative return events, photons are radiated along the
direction of the incoming beam and lost in the beam pipe.
Only about 10\% to 15\% of photons are reconstructed in the ILD
detector in the best case \cite{Kalinowski:2021tyr}.
Because of the multi-photon radiation and ISR contributions included
in the presented study simple selection of events based on the narrow
window cuts in photon and di-muon energies, as suggested in the FCC-ee
study \cite{GonzalezSuarez:2024dsp}, also turned out not to be very efficient. 
That is why we decided to look for alternative signal selection criteria,
not relying on the photon reconstruction but based on the expected
kinematics of radiative events. 

\begin{figure}[tbp]
    \centering
    \includegraphics[width=0.45\textwidth]{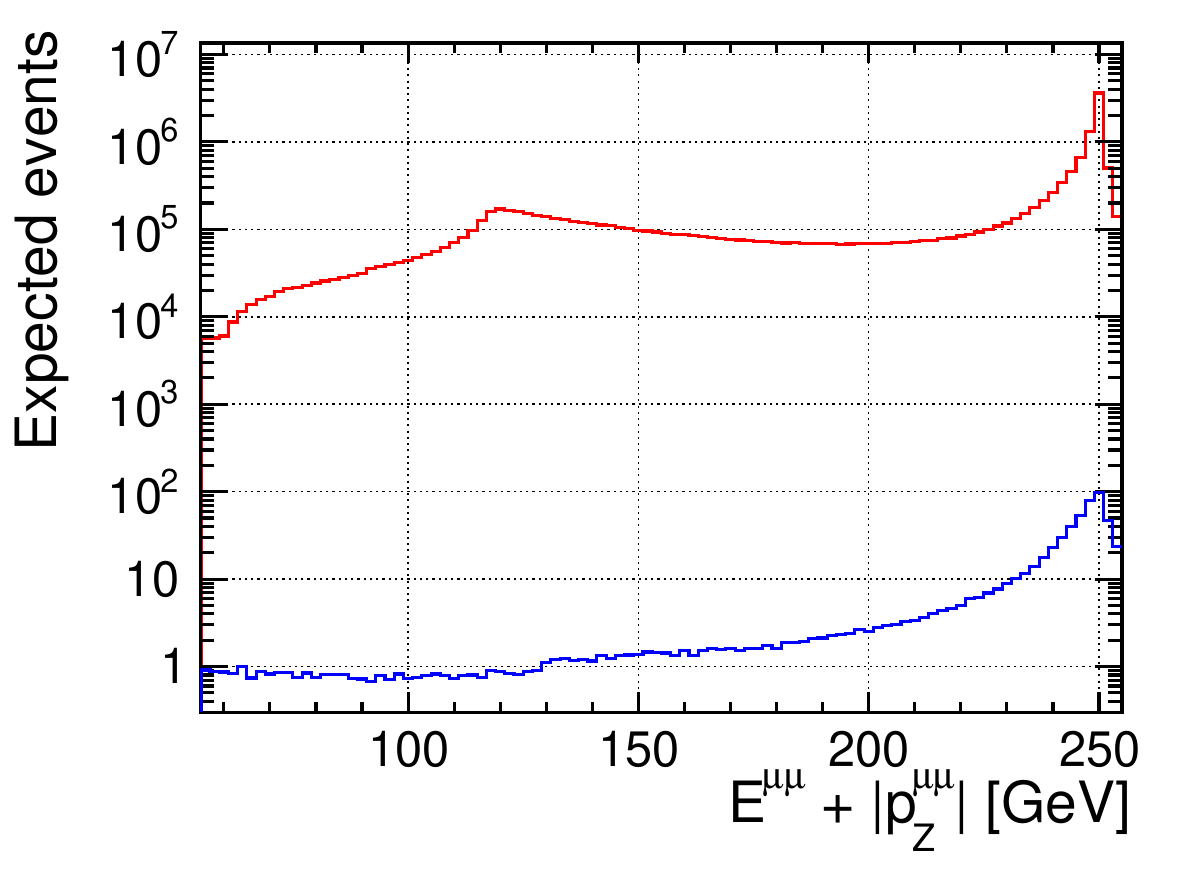}
    \includegraphics[width=0.45\textwidth]{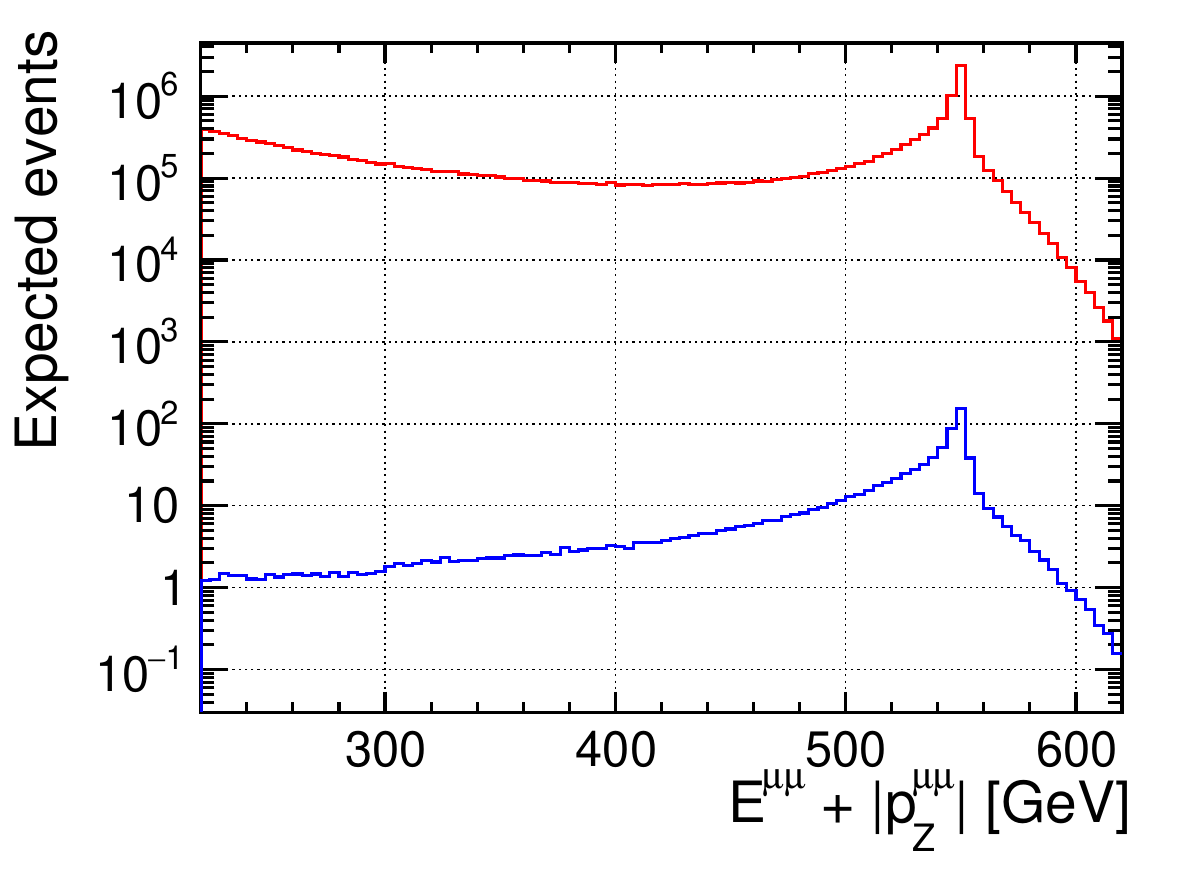}
    \caption{Distribution of the sum of the reconstructed muon pair
      energy and the absolute value of its longitudinal momentum for
      SM background events (red histogram) and the Z' production
      events (blue histogram). 
      Left: for ILC running at 250\,GeV and Z' mass of 50\,GeV.
      Right: for LCF running at 550\,GeV and Z' mass of 200\,GeV.
      Signal samples are normalised to Z' coupling of $g_{Z'}=0.001$.
}  
    \label{fig:epz}
\end{figure}
\begin{figure}[tbp]
    \centering
    \includegraphics[width=0.45\textwidth]{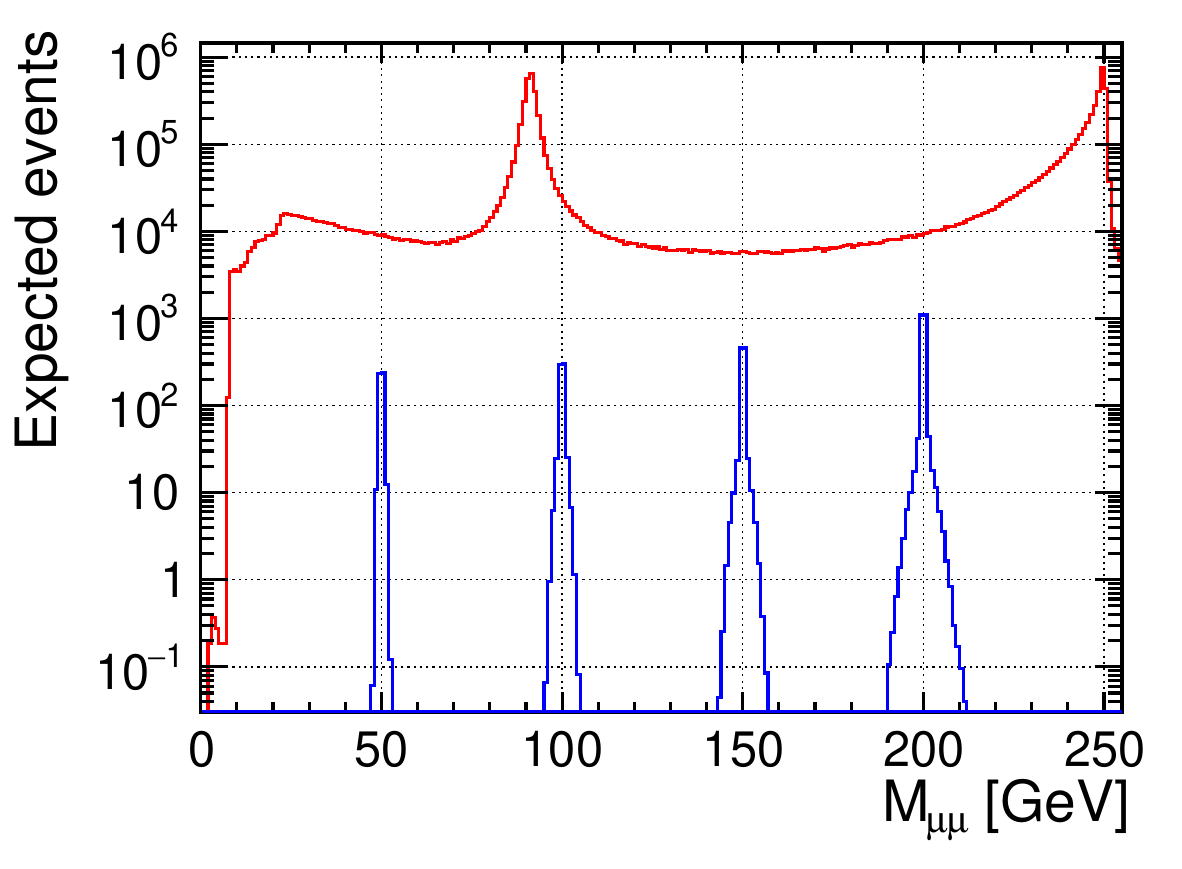}
    \includegraphics[width=0.45\textwidth]{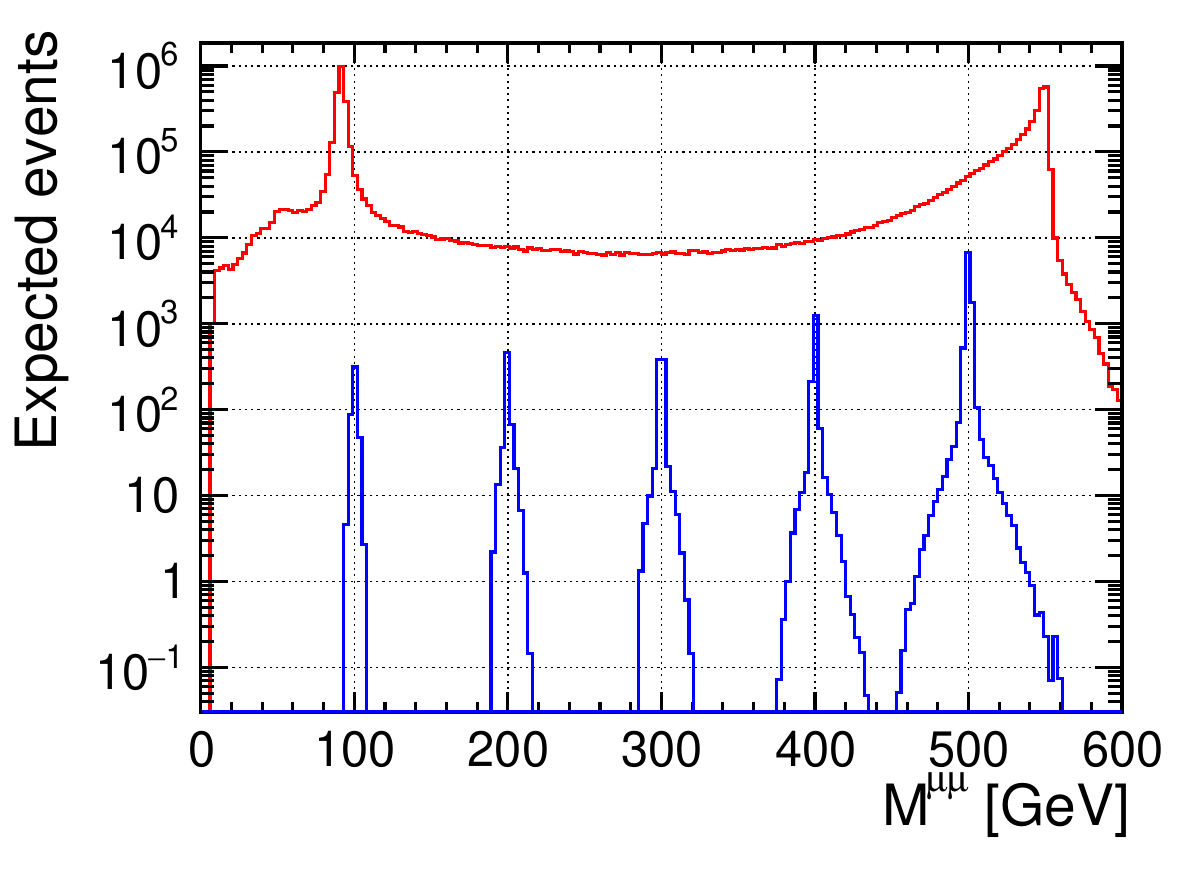}
    \caption{Distribution of the reconstructed muon pair invariant
      mass for SM background events (red histogram) and the Z' production events,
      after radiative event selection cuts (blue histograms).
      Left: for ILC running at 250\,GeV and Z' mass of 50\,GeV,
      100\,GeV, 150\,GeV and 200\,GeV.
      Right: for LCF running at 550\,GeV and Z' mass of 100\,GeV,
      200\,GeV, 300\,GeV, 400\,GeV and 500\,GeV.
      Signal samples are normalised to Z' coupling of $g_{Z'}=0.001$.
    }  
    \label{fig:minv}
\end{figure}

For radiative Z'-return events, the momentum of the radiated photon
should be balanced by the momentum of the produced on-shell Z'. 
The value of the longitudinal momentum of the di-muon system,
$|P^{(\mu\mu)}_Z|$ can be thus used as an estimate of the momentum
(and thus also energy) of the photon radiated along the beam direction,
even if photon itself is lost in the beam pipe.
This allows for efficient selection of radiative di-muon events with a
cut on the energy and longitudinal momentum sum, $E^{(\mu\mu)} +
|P^{(\mu\mu)}_Z|$.
Distributions of this variable for signal and background events, for
ILC running at 250\,GeV and LCF running at 550\,GeV are shown in
Fig.~\ref{fig:epz}. 
With an additional cut on the total transverse momentum of the event,
$P^{(tot)}_T$, suppressing the contribution of processes with
neutrinos in the final state the dominant
background contribution comes from the SM production of di-muon pairs,
$\ee \to \mumu (\gamma)$. 
Distributions of the reconstructed muon pair invariant mass for SM
background events and for the Z' production events, for selected boson
masses, after cuts applied on $E^{(\mu\mu)} + |P^{(\mu\mu)}_Z|$ and
$P^{(tot)}_T$, are compared in Fig.~\ref{fig:minv}.
As the dominant background is due to the SM muon-pair production (with
$\gamma/Z^0$ exchange), additional selection cuts based on the photon
reconstruction cannot improve the sensitivity, as the spectra of radiated photons is
expected to be the same for the considered signal and the background.  

\section{Results}

\begin{figure}[tbp]
    \centering
    \includegraphics[width=0.45\textwidth]{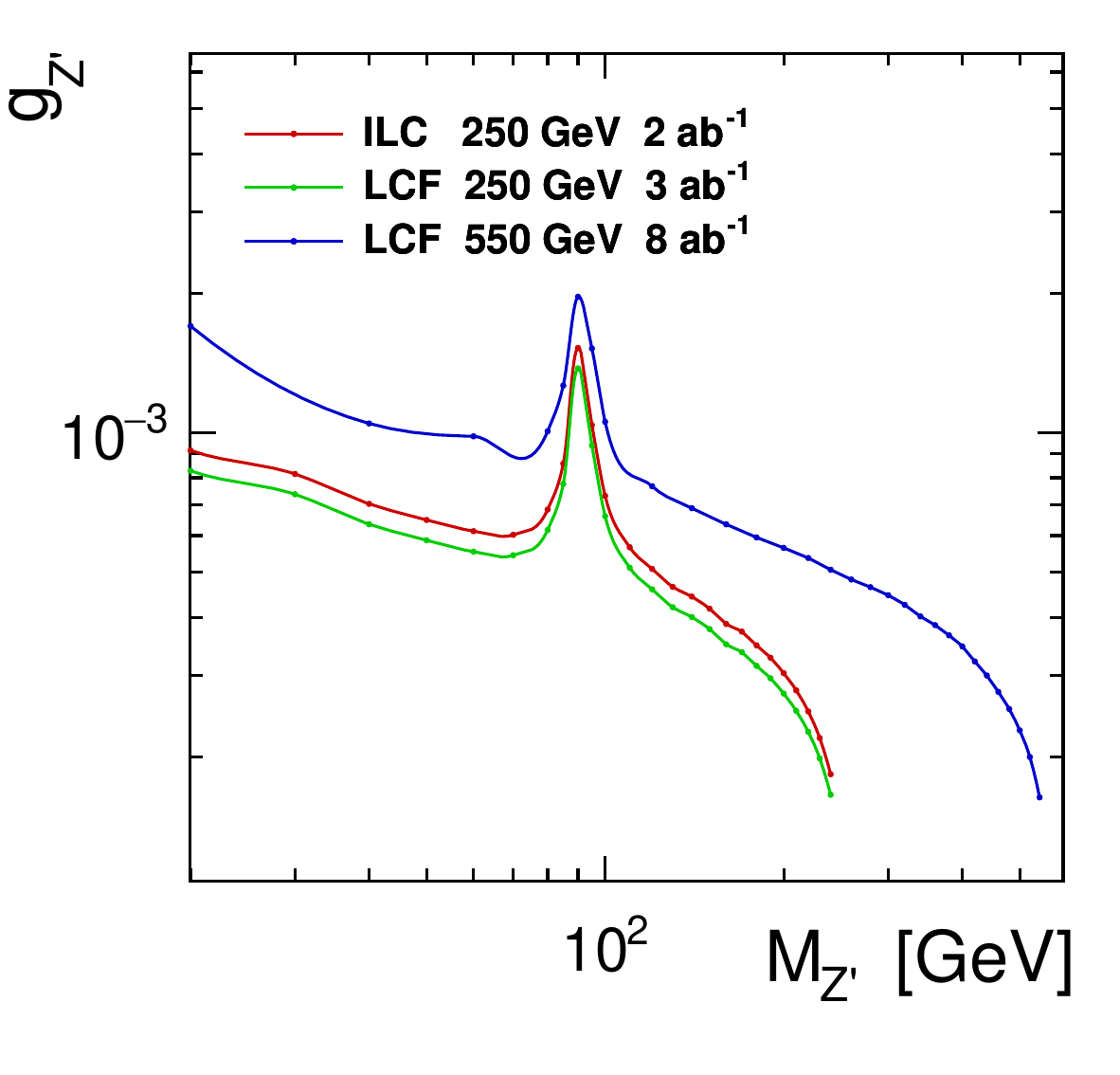}
    \includegraphics[width=0.45\textwidth,trim=0 -0.45cm 0 0.45cm]{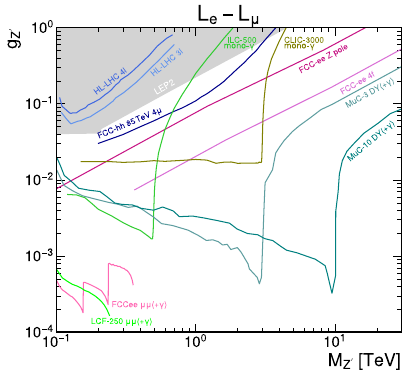}
    \caption{Expected \SI{95}{\%} CL exclusion limits on the leptophilic
      Z' boson coupling. Left: limits expected from the di-muon
      invariant mass distribution for $\ee \to \mumu (\gamma)$
      events,  based on fast detector simulation for for the
      ILC and LCF running at \SI{250}{GeV}, and LCF running at
      \SI{550}{GeV}. Right: comparison of LCF limits at \SI{250}{GeV}
      with existing limits from LEP2 and projected exclusion curves
      for HL-LHC and other future facilities (figure taken
      from~\cite{deBlas:2025gyz}).}  
    \label{fig:zprime}
\end{figure}

In the coupling value range considered in this study, the intrinsic
width of the Z' is expected to be in ${\cal O}$(\SI{10}{keV}) range,
negligible compared to the experimental resolution.
For LCF running at \SI{250}{GeV}, the invariant mass of the di-muon
from the Z' decay can be reconstructed with a resolution of up to
\SI{0.7}{GeV}, which worsens to about \SI{2}{GeV} for LCF running at
\SI{550}{GeV} due to the higher momenta of the produced muons.
The signal of Z' production should be clearly visible as a narrow
resonant peak in the di-muon invariant mass distribution on top of the
smooth, structure-less (except for the peaks at the Z-mass and at
the full collision energy) SM background.
Expected limits on the coupling of the Z' boson to SM leptons,
assuming no deviations from the SM predictions are observed, are
extracted from the template fit to the invariant mass distribution, see
Fig.~\ref{fig:minv}.
Results are presented in Fig.~\ref{fig:zprime} (left) for ILC and LCF running at
\SI{250}{GeV}, with a total integrated luminosity of \SI{2}{\abinv}
and \SI{3}{\abinv}, respectively, and LCF at \SI{550}{GeV}
with \SI{8}{\abinv}.
Luminosity sharing for ILC and LCF running with different beam polarisation
settings is taken from \cite{LinearColliderVision:2025hlt}. 
Assumed is the electron beam polarisation of 80\% and positron
polarisation of 30\% and 60\% for 250\,GeV and 550\,GeV running, 
respectively.
Also included in  Fig.~\ref{fig:zprime} (right) is the comparison of
the expected limits from LCF at \SI{250}{GeV} with current LEP2
constraints, HL-LHC projections and prospects for other future
facilities, as presented in \cite{deBlas:2025gyz}.

Despite the lower integrated luminosity, the
limits for LCF at \SI{250}{GeV} are better than those for FCC-ee at
\SI{240}{GeV}.
Multiple factors contribute to this effects. Some are not intrinsic to
the collider properties, but to the analysis, i.e.\ the benefit from a
higher signal cross-section estimate with multi-photon emission
included and the improved selection strategy not requiring a
reconstructed photon in the final state.
Others, however, are intrinsic to the choice of collider, in
particular the beam polarisation, helping with signal-to-background
ratio, as well as the better muon momentum resolution resulting from
higher magnetic field of detectors at linear colliders, which is not
possible at circular colliders as it would have too destructive
effects on the circulating beam.  

\section{Conclusions}

Presented in this contribution is the fast simulation analysis of
leptophilic Z' search prospects for ILC and LCF running at 250\,GeV
and LCF running at 550\,GeV.
The analysis is based on detailed event generation in \whizard, with
effects of ISR and multi-photon emission properly included for signal 
and background.
A novel method of radiative event selection based on muon-pair
kinematics is introduced, significantly increasing signal selection
efficiency. 
Expected limits on the Z' couplings, based on the reconstructed di-muon invariant
mass distribution,  are slightly better than those presented in the similar FCC-ee study.
While systematic effects were not considered in the study, they are
not expected to be significant in the search for very narrow
resonance.

\subsection*{Acknowledgments}

This work, carried out in the framework of the ILD concept group, 
was supported by the National Science Centre,
Poland, under the OPUS research project no. 2021/43/B/ST2/01778.

\printbibliography

\end{document}